\documentclass[10pt,letterpaper]{article}

\usepackage{graphicx}                       
\usepackage{geometry}                       
\usepackage{amsmath,amsfonts,bm,amssymb}            
\usepackage{enumitem}                       
\usepackage{natbib}                         
\usepackage[hidelinks, colorlinks=true, 
linkcolor=black, citecolor=blue]{hyperref}  
\usepackage{fancyhdr, calc, lastpage}       
\usepackage{booktabs}                       
\usepackage{lipsum}                         
\usepackage{endnotes}                       

\fancyheadoffset[L,R]{\marginparsep+1cm} 
\fancyfootoffset[L]{\marginparsep+1cm} 
\newcommand\setheader[2]{
    \fancyhead[L]{\footnotesize #1}
}   

\renewcommand\title[1]{{\linespread{1} \noindent\LARGE \bf \hskip2.25pc \parbox{.8\textwidth}{%
\LARGE \bf \begin{center} #1 \end{center}\rm } \rm\normalfont\normalsize} }
\renewcommand\author[1]{{\linespread{1} \noindent\hskip2.25pc \parbox{.8\textwidth}{%
   \normalsize \bf \begin{center} #1 \end{center}\rm } \vskip-1.4pc }}
\newcommand\address[1]{{\linespread{1} \noindent\hskip2.25pc \parbox{.8\textwidth}{%
   \footnotesize \it \begin{center} #1 \end{center}\rm }  \normalsize \vskip-1pc }}

\newcommand\PACS[1]{\vskip-2.75pc \begin{center}\parbox{.8\textwidth}{\small\bf PACS numbers: \rm #1 \hfill} \end{center}\vskip4pt}%

\renewenvironment{abstract}
{\vskip1pc\noindent\begin{center} \begin{minipage}{.8\textwidth} {\bf Abstract: } }
{ \vspace{.25cm} \end{minipage}\end{center}\normalsize\vskip-1.5pc}%

\def\fps@table{h}
\renewcommand\refname{\normalsize References and links \rm}

\makeatletter
\newcommand\@MaxCapWidth{5.2in}
\long\def\@makecaption#1#2{%
  \small
  \vskip\abovecaptionskip
  \sbox\@tempboxa{#1. #2}%
  \ifdim \wd\@tempboxa >\@MaxCapWidth
    \parbox{5.2in}{#1. #2}
  \else
    \global \@minipagefalse
    \hb@xt@\hsize{\hfil\box\@tempboxa\hfil}%
  \fi
  \vskip\belowcaptionskip\normalsize}
\makeatother

\makeatletter
\renewcommand\@seccntformat[1]{\csname the#1\endcsname.\hspace{.1cm}}

\renewcommand\section{\@startsection {section}{1}{0pt}%
                                     {-2ex plus -1ex minus -.2ex}%
                                     {0.65ex plus 1.2ex}%
                                     {\normalsize\bfseries}}
\renewcommand\subsection{\@startsection{subsection}{2}{0pt}%
                                     {-2.25ex plus -1ex minus -.2ex}%
                                     {.45ex plus .2ex}%
                                     {\normalsize\itshape}}
\renewcommand\subsubsection{\@startsection{subsubsection}{3}{0pt}%
                                     {-2.25ex plus -1ex minus -.2ex}%
                                     {1ex plus .2ex}%
                                     {\small\upshape}}
\makeatother

\makeatletter
\let\old@theendnotes\theendnotes
\renewcommand{\theendnotes}{\old@theendnotes\vspace{.3cm}}
\makeatother
\let\footnote=\endnote 

\makeatletter
\renewenvironment{thebibliography}[1]
     {\section*{\refname}%
      \@mkboth{\MakeUppercase\refname}{\MakeUppercase\refname}%
        \footnotesize

        \ifnum\value{endnote} > 0
        \theendnotes 
        \fi

      \list{\@biblabel{\@arabic\c@enumiv}}%
           {\settowidth\labelwidth{\@biblabel{#1}}%
            \setlength\itemindent{0pt}
            \setlength\itemsep{-1pt}
            }}
     {\endlist}
    
    \expandafter\ifx\csname natexlab\endcsname\relax\fi
    \expandafter\ifx\csname url\endcsname\relax
    \def\url#1{\texttt{#1}}\fi
    \expandafter\ifx\csname urlprefix\endcsname\relax\fi
    \providecommand{\bibinfo}[2]{#2}
    \providecommand{\noopsort}[1]{}
    
\makeatother

\begin{document}

\setheader{In situ monitoring of plastic deformation}{Salinas}

\title{In situ monitoring of plastic deformation using ultrasound}

\author{Vicente Salinas, Fernando Lund, Nicol\'as Mujica}
\address{Departamento de F\'isica, Facultad de Ciencias F\'isicas y Matem\'aticas, Universidad de Chile, Avenida Blanco Encalada 2008, Santiago, Chile}
\author{Rodrigo Espinoza-Gonz\'alez}
\address{Departamento de Ciencia de los Materiales, Facultad de Ciencias F\'isicas y Matem\'aticas, Universidad de Chile, Avenida Tupper 2069, Santiago, Chile}


\begin{abstract}

{Ultrasound has long been used as a  non-destructive tool to test for the brittle fracture of materials. Could it be used as a similar tool to test for ductile failure? This study reports results of local measurements of the speed of shear waves, $v_T$, in aluminum under standard testing conditions at two different locations on the same sample, as a function of stress. There is a clear change in $v_T$ at the Yield stress, consistent with a proliferation of dislocations.  This measurement provides a quantitative, continuous relation between dislocation density and externally applied stress.}

\end{abstract}

\section{Introduction}

Ultrasound (US) has been in use for decades as a nondestructive testing tool \citep{McSkimin1960,chen2007}. The reason is that, because of the low energies involved,  it can penetrate deep into a material without affecting it. The detection of cracks and flaws in solid materials in service is a major field of application. 

The propagation of cracks causes brittle fracture and one concern is to detect them before they reach a critical size for catastrophic propagation. Another mode of failure is ductile failure or plastic yield, which is governed by the proliferation of dislocations. The latter are line defects in crystalline solids that explain why the experimental value of the shear stress needed to plastically deform a crystal is several orders of magnitude less than the theoretical value, obtained on the basis of the shear stress needed to rigidly slide one atomic plane past an adjacent one. Can US be used as a nondestructive testing tool for the plastic behavior of materials in the same way that it is used to test for brittle fracture?

For a long time it has been known that US, or, more generally, elastic waves, interact with dislocations  \citep[and references therein]{maureletal}. However, the interaction was always thought to be too weak to yield a useful signal. Recently, however, \citet{Mujica2012} showed, using resonant Ultrasound Spectroscopy (RUS) \citep{migliori93,leisure97,Ogi2002}, that metals often have dislocations in high enough numbers that their collective behavior does provide a measurable signal. More specifically, they showed that dislocation densities in aluminum at the $10^9$~mm$^{-2}$ level produced a shift in the speed of propagation of shear waves at the $1\%$ level. These results were obtained with samples especially prepared for laboratory testing. Can this technique be scaled to pieces in service?

One intermediate step that must be taken is to see whether the dislocations that are generated by metals and alloys under standard testing conditions generate a shift in wave speed propagation that can be measured with available hardware. This paper presents results of an experiment that answers this question in the affirmative. {In previous work with similar conditions, \citet{hiraoetal00} used the method of electromagnetic acoustic resonance (EMAR) to make a contactless average masurement of ultrasonic attenuation and velocity during fatigue tests of polycrystalline copper. \citet{minkato04} monitored the ultrasonic parameters of an aluminum alloy inside a water bag under cyclic loading and \citet{minetal} used a similar apparatus to measure the speed of sound of the same material during tensile testing. }

\section{Experimental setup and procedures}

\begin{figure}[t!]
\centering
	\includegraphics[height=4.8cm]{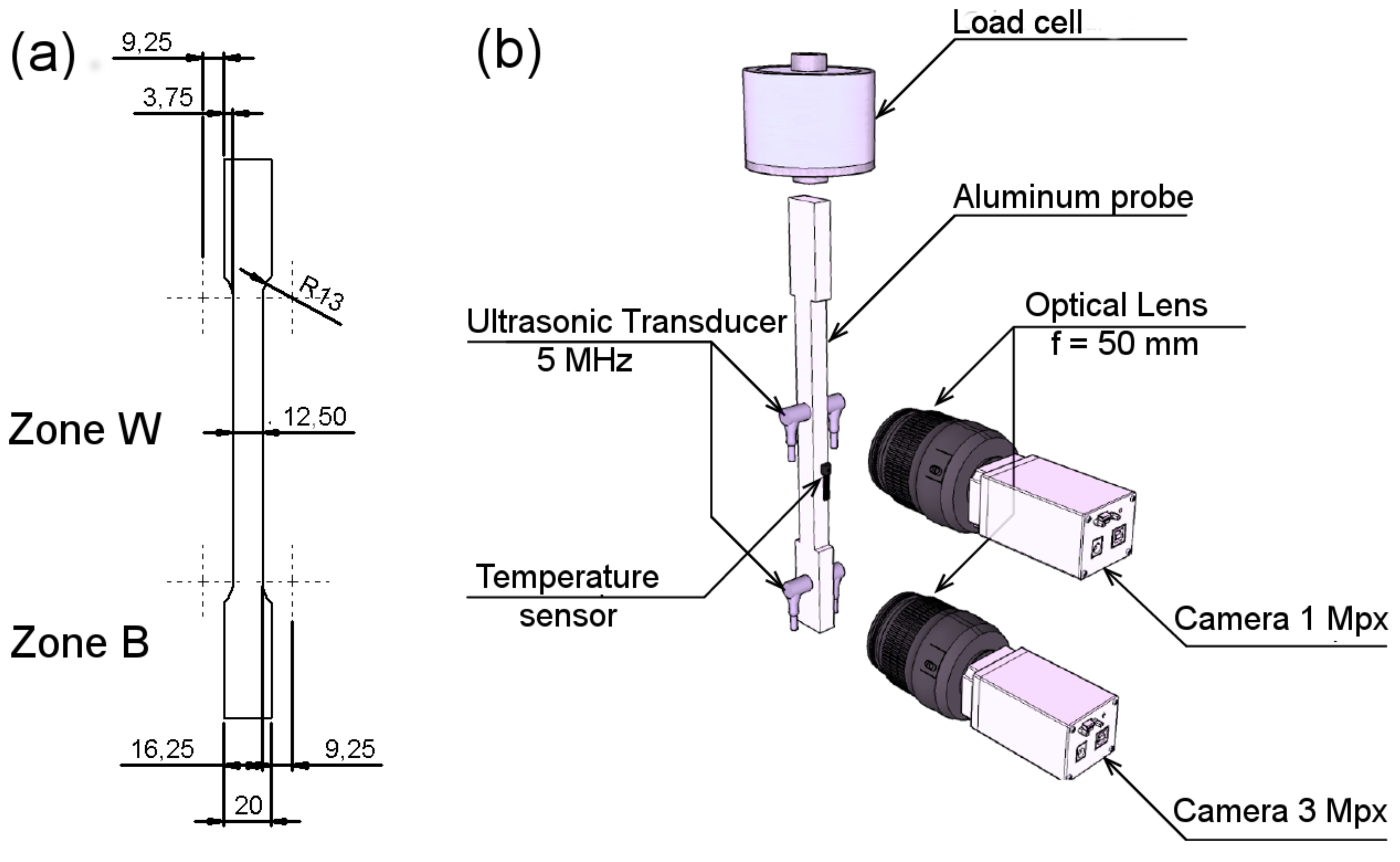}
		\label{fig:esquema_montaje}
			\caption{(Color online) (a) Aluminum probe dimensions under standard ASTM E8/E8M. Samples are $240$ mm long and have a rectangular cross section, with symmetric top and bottom parts, $50$ mm long, that have a larger cross area ($10\times20$ mm$^2$). The central part, about $120$ mm long, has a smaller cross area ($10\times12.5$ mm$^2$). The central part and the symmetric top and bottom sections are separated by transition zones $9.13$ mm long.  (b) Schematic of the experimental setup. Two pairs of ultrasonic transducers are fixed to the sample, one at zone B and the other at zone W. Each pair of transducers is held with a constant compression force using a holder and a system of springs (not shown in the figure for simplicity). The local width at each zone is measured by images acquired with digital cameras.}
\end{figure}

Figure \ref{fig:esquema_montaje} presents details of the probes and a schematic drawing of the experimental setup. The probes were fabricated out of aluminum with a geometry that is defined in the standard ASTM-E8/E8M for tensile tests \citep{ASTM}. Here, we present results obtained with two samples, which were cut by Electrical Discharge Machining (EDM), which we name P1, P2. Consequently, the parallelism of opposite faces is better than $0.3^\circ$, which results in variations of width  that are smaller than $60$~$\mu$m over a distance of $12.5$ mm. Details are given in Fig. \ref{fig:esquema_montaje}(a). 

Tensile tests are performed with an Instron 3369 machine. The maximum load capacity is $50$ kN, which implies a maximum stress of $250$ MPa in the large cross section areas and of $400$ MPa in the smaller cross section central part. These values are well suited in the range of aluminum yield stress and ultimate stress, implying that our samples can indeed fracture. Special grips from Instron are used due to the ductile nature of aluminum in order to minimize sliding. The tensile tests are performed at an initial speed of $0.01$ mm/min for loads lower than $3.5$ kN (elastic regime) and later at $0.03$ mm/min for larger loads (plastic regime). The tests are then considered as quasi-static, but at the same time have a maximum duration of $10$ h, which cannot be much longer for operation conditions.  For some tests, the probes were compressed back, at a speed of $0.03$ mm/min, until the applied load was null.

In Fig. \ref{fig:esquema_montaje}(b) we present a schematic representation of the experimental setup, including the most important elements. Two pairs of transverse wave ultrasonic transducers (Olympus V157-RM) are placed in contact with the probe using an ultrasonic couplant. One pair is placed at the base (zone B) and the other at the center of the probe (zone W), each corresponding to the large and small cross sectional area respectively. The ultrasonic transducers are centered at $5.49$ MHz and have a wide frequency response (of $5.5$ MHz for $-6$ dB). Additionally, two strain gages (Omega KFH-3-350-C1-11L1M2R) are placed in the vicinity of each pair of ultrasonic transducers in order to measure the local strain. The strain gages are $3$~mm long and their maximum deformation is $3\%$. Their voltage signals are measured with a Wheatstone bridge in configuration quarter-bridge and acquired by a analog to digital converter (National Instruments NI PCI-6381), which insure a voltage resolution of $3.8$~$\mu$V, thus a precision of $0.635$ $\mu$m/m. Finally, an integrated-circuit temperature sensor (Texas Instruments LM35) is placed between the two ultrasonic transducer pairs allowing a precise temperature measurement ($0.01$~$^\circ$C).

The transverse elastic wave speed is then measured locally in both zones B and W. The main goal is to measure wave speed variations during plastic deformation, which should be larger in zone W because of its reduced area and thus larger local strains. The wave speeds are determined by measuring the time of flight of short ultrasonic pulses. 
These consist of carrier signals with frequency $5$ MHz modulated by a Gaussian curve, resulting in a short pulses of $5$ cycles. The carrier's wavelength is $\lambda = 0.62$ mm and the total pulse length is $\approx 3.1$ mm, which indeed is short compared to the pulse propagation distance ($10$ mm). The excitation signal is generated by a Labview program that is sent to a waveform generator (Agilent 33500B) by means of a General Purpose Interface Board (NI GPIB-USB-HS). This excitation signal is then amplified by a high-power bipolar amplifier (NF Techno Commerce BA4850). Both the excitation voltage signal and the measured received signal are acquired by a high speed oscilloscope (Lecroy Wavejet 334A). All the ultrasonic signals are then transferred to a personal computer (PC) with the same GPIB protocol. The time of flight is then obtained by computing the cross correlation between the excitation and received signal envelopes, which in turn are computed as the absolute values of their Hilbert transforms. The final time resolution is $1$ ns.

During the tensile tests the probe's cross sectional area decreases. In particular, each transverse dimension decreases. In the elastic regime this is quantified by the Poisson ratio $\nu$. Because the wave speed is determined as the ratio between the time of flight and the probe width, we need to quantify these variations. This is done locally at each zone B and W by means of two digital cameras, of resolutions $3$~MP and $1$~MP respectively. As variations in the lower strain region (zone B) are expected to be smaller, the camera with the better resolution is used. The final spatial resolutions are $5.7$ $\mu$m/pixel and $8.5$ $\mu$m/pixel for zones B and W respectively.

The tensile testing machine provides the measurement of the total load during the probe's deformation. Additionally, we measure synchronously all the other physical relevant quantities, such as the emitted and received signals, the local strains, the sample's temperature and the images of each camera that capture the local width at each transducer pair location. This is done by means of an analog trigger connector included in the tensile testing apparatus, which triggers the data and image acquisition from the PC, which runs a in-house developed Labview program.

Temperature variations were kept small during the experimental runs, with a maximum variation of about $3$ $^\circ$C. There are two important reasons to do so: (i) the strain versus voltage measurements are temperature dependent, which is measured by a proper calibration. Thus, the real strain is computed by making this temperature correction. (ii) The sample also varies its volume by thermal expansion. This can not be detected by the image analysis.  Simultaneous time of flight and temperature measurements with no load applied to the samples allow us to calibrate and correct this effect. For both corrections, the data are computed with the reference temperature $T = 25$~$^\circ$C. 

\begin{figure}[t!]
\centering
	\includegraphics[height=4.5cm]{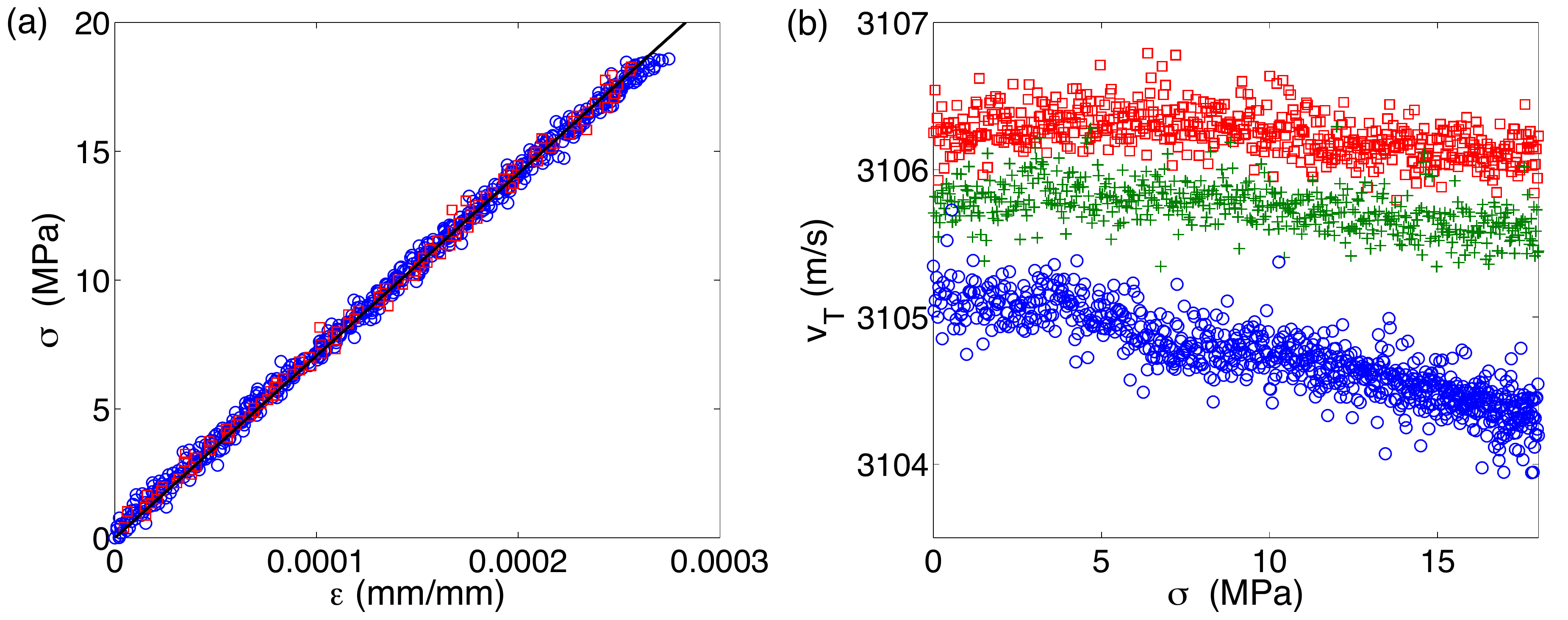}
		\caption{(Color online) (a) Third elastic tensile ({$\circ$}) and compression ({$\square$}) tests on sample P2 in zone B. For zone W the results are qualitatively very similar. The data is fitted to $\sigma = E \varepsilon$ (continuous line), where $E$ is the Young's modulus. We obtain $E = 70.7\pm 0.2$~GPa for zone B and $E = 72.2\pm 0.2$~GPa~for zone W. The $R^2$ regression coefficients are $0.996$ and $0.997$ respectively. (b)~$v_T$ versus $\sigma$ during the first ({$\circ$}), second ({$+$}) and third ({$\square$}) elastic tensile tests (zone B).}
	\label{fig:ensayo_elastico}
\end{figure}

\section{Results}
\subsection{Elastic regime}

Before plastic deformations are performed we realize three elastic deformations. The main reason is that we need to verify that the elastic wave speed does not vary, or varies very little, for deformations within the elastic regime. Additionally, we need to allow the glued strain gages to ``settle" well on the probe, as well as the testing machine grips. An example of the third elastic tensile and consecutive compression tests are shown in Fig. \ref{fig:ensayo_elastico}(a) for probe P2 with data obtained from zone B. These are reversible curves, independent of the deformation speed. For zone W the results are qualitatively very similar, except that the local stress increases up to $\sim30$ MPa.

The measured transverse elastic wave speeds vary less than $1$ m/s during the complete elastic test (For an example see Fig. \ref{fig:ensayo_elastico}(b)). During the three elastic consecutive tests it either varies randomly around a mean value or it shows a slow systematic trend to either increase or decrease but always less than $1$ m/s. For zone B, the measured wave speeds during the three consecutive tensile deformations, from $0$ MPa to $19$ MPa, are $v_T = 3104.7 \pm 0.3$ m/s (test 1), $v_T = 3105.7 \pm 0.2$ m/s (test 2) and $v_T = 3106.2 \pm 0.2$ m/s (test 3). For zone W, the measured wave speeds, varying the stress from $0$ MPa to $30$ MPa, are $v_T = 3116.7 \pm 0.2$ m/s (test 1), $v_T = 3118.4 \pm 0.2$ m/s (test 2) and $v_T = 3119.0 \pm 0.3$ m/s (test 3). At first approximation, these values are all constant. However, small variations might be due to slight re-accommodation of the ultrasonic transducers, which might get more o less well aligned during the tensile and compression tests because during the deformation a small rotation of the probe around its principal axis certainly occurs. The differences between zone B and W are also attributed to the camera's different spatial resolutions. We have done many measurements with no load with different placements of the transducer pairs and the estimated error of reproducibility, due to slightly different alignments, is of the order of $1-2$ m/s. So, changes of this order of magnitude are attributed to changes of transducer accommodations on the probes. 

\begin{figure}[t!]
\centering
	\includegraphics[height=4.5cm]{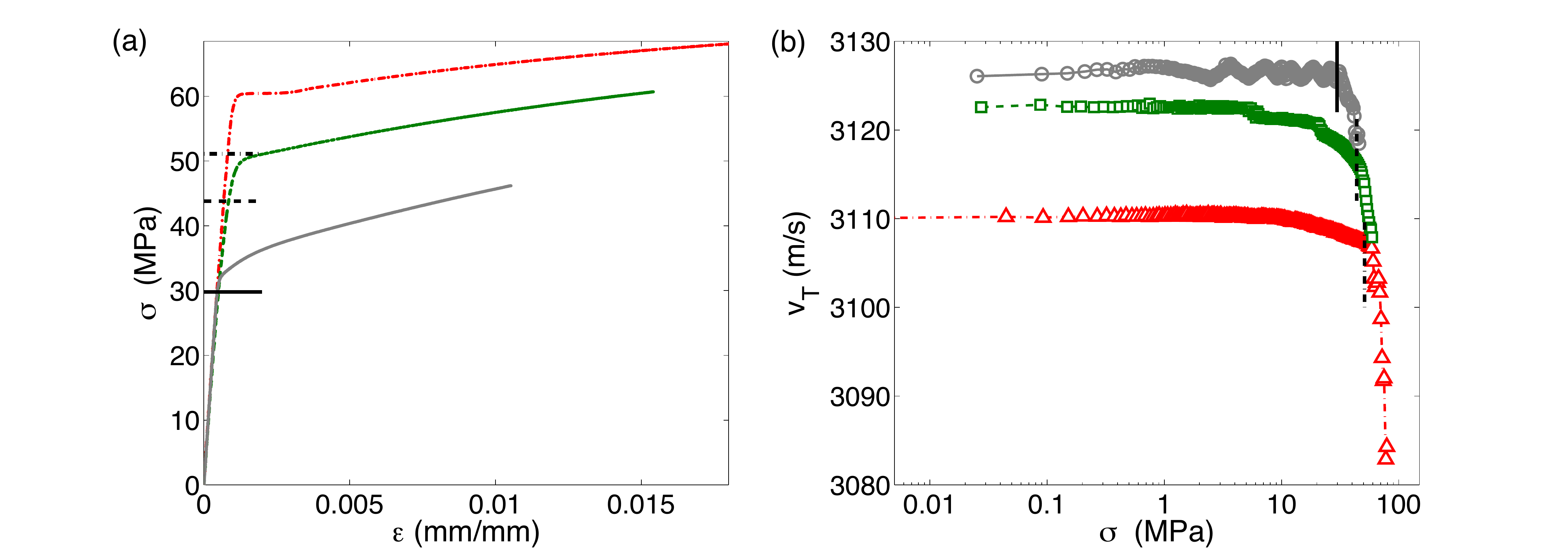}
	\caption{(Color online) (a) Stress-strain curves for three consecutive tensile tests for sample P1 (zone W). The  Yield stress increases for each consecutive test. (b) $v_T$ versus $\sigma$ in semi-log scale for the same tensile tests (first: {$\circ$}, second: {$\square$} and third test: {$\triangle$}). Each new test starts with a wave speed that is very close to the final value of the previous test. The wave speed decreases abruptly at $\sigma \sim \sigma_Y$ and all curves colapse in the plastic regime. The horizontal and vertical lines, in (a) and (b) respectively, correspond to the measured Yield stress following \citet{Christensen2008}.}
	\label{fig3}
\end{figure}
\subsection{Plastic regime}

We now turn to the results obtained within the plastic regime. In Fig. \ref{fig3}(a) we show the stress versus strain curves obtained from zone W for three consecutive tensile tests, each one finishing at a larger maximum stress. The yield stress $\sigma_Y$ increases with the number of tests, which is a well known hardening induced by the proliferation of dislocations \citep[Chap. 5]{Reedhill1992}. The corresponding Yield stress's, computed by the method proposed by \citet{Christensen2008}, are $29.8\pm0.4$ MPa, $43.8\pm0.3$ MPa, $51.1\pm0.5$ MPa respectively. We emphasize that this method uses the data from the stress versus strain curve, specifically by imposing $(d^3 \sigma/d\varepsilon^3)|_{\varepsilon_Y} = 0$ and then computing $\sigma_Y = \sigma(\varepsilon_Y)$. This definition considers the point $(\varepsilon_Y,\sigma_Y)$ for which the tangent modulus is changing at the fastest rate for increasing strain. In Fig. \ref{fig3}(b) we present the corresponding transverse elastic wave speeds measured during these same deformation tests. Here, the data is presented in semi-log scale to emphasize the abrupt changes that occur for $\sigma > \sigma_Y$. For each new test, the initial wave speed is very close to the final wave speed at the end of the previous test. Additionally, for $\sigma>\sigma_Y$, the measurements tend to collapse on a single curve regardless of the initial condition. 

\begin{figure}[t!]
\centering
	\includegraphics[height=4.5cm]{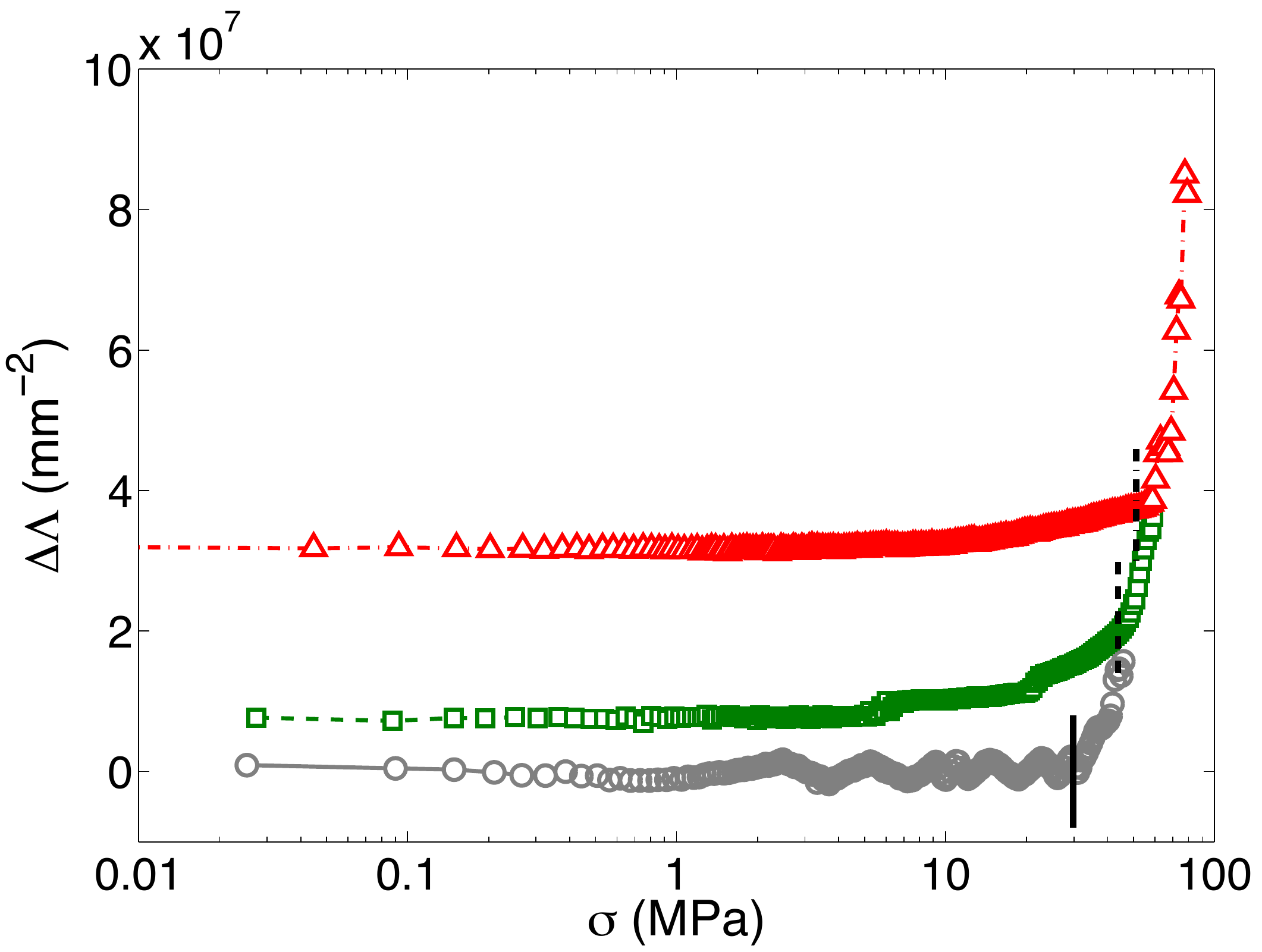}
	\caption{(Color online) $\Delta\Lambda$ versus $\sigma$ in semi-log scale, for zone W and $L=10$ nm (sample P1). The data is from the same tensile tests shown in Fig. \ref{fig3} (first: {$\circ$}, second: {$\square$} and third test: {$\triangle$}). Vertical lines indicate the measured $\sigma_Y$. For $L=100$ nm the results are quantitatively the same but divided by a factor $100$.}
	\label{fig4}
\end{figure}

Using the analysis reported in \citet{Mujica2012}, we can estimate changes in surface dislocation density $\Lambda$ using the expression $\Delta v_T/v_T^0= -8\Delta(\Lambda L^2)/(5\pi^4)$, where $L$ is the dislocation length. Here, we use $\Delta v_T = v_T - v_T^0$, where $v_T$ is the measured wave speed and $v_T^0$ is the initial value, before the first plastic deformation.  In Fig. \ref{fig4} we show our results for the measurements of zone W and assuming $L = 10$~nm. The curves remain quantitatively identical if we use $L=100$ nm, except that the number of dislocations per unit area decreases by a factor of $100$.  Notice that this is a continuous monitoring of dislocation density as a function of stress, a measurement that would be impossible to perform either with X-ray diffraction (XRD) or Transmission Electron Microscopy (TEM), given their intrusive nature. We obtain changes in the range $\Delta \Lambda \sim 1\times 10^7 - 1\times 10^8$ mm$^{-2}$, which is very reasonable given the small strains that are imposed. We expect that for these small applied strains, any additional influences, such as a change in grain size, would be negligible. This figure also demonstrates that abrupt changes occur at $\sigma\sim \sigma_Y$ and that the curves collapse above this value. 

\section{Conclusions}
The main result of this communication is encapsulated in Figs. \ref{fig3} and \ref{fig4}: The onset of plasticity in aluminum, as determined by the Yield stress, is accompanied by a marked decrease in the speed of shear waves, consistent with a proliferation of dislocations. The measurement of the speed of shear waves as a function of stress enables a quantitative, continuous, relation between dislocation density and externally applied stress. The traditional methods to measure dislocation density, such as XRD or TEM, are not able to provide such a continuous link because their intrusive nature only allows for before-and-after measurements.

\section*{Acknowledgments}
We acknowledge the support of Grant Fondecyt 1130382 and Conicyt Grant ANR 38 - PROCOMEDIA.



\begin{thebibliography}{1}    



\bibitem[{ASTM International(2013)}]{ASTM}  
\bibinfo{editor}{ASTM International}
  (\textbf{\bibinfo{year}{2013}}). {\bibinfo{title}{``ASTM E8 / E8M-13a, Standard Test Methods for Tension Testing of Metallic Materials"}},  {\bibinfo{subtitle}{\it ASTM International, West Conshohocken, PA}}, \bibinfo{http}{http://www.astm.org}.
  
 \bibitem[{Chen(2007)}]{chen2007}
 \bibinfo{editor}{Chen, C.-H.}
 (\textbf{\bibinfo{year}{2007}}). 
 {\bibinfo{booktitle}{``Ultrasonic and Advanced Methods for Nondestructive Testing and Material Characterization"}}, 
 \bibinfo{publisher}{(World Scientific, Singapore)}.
 
 
 \bibitem[{Christensen(2008)}]{Christensen2008}
 \bibinfo{author}{Christensen, R.M.}
 (\textbf{\bibinfo{year}{2008}}). 
 {\bibinfo{title}{``Observations on the definition of yield stress"}}, 
 \bibinfo{journal}{Acta Mech.} 
 \textbf{\bibinfo{volume}{196}}, 
 \bibinfo{pages}{ 239--244}.
 
\bibitem[{Hirao et al.(2000)}]{hiraoetal00}
 \bibinfo{author}{Hirao, M., Ogi, H., Suzuki, N., Ohtani, T. and Kato, H.}
 (\textbf{\bibinfo{year}{2000}}). 
 {\bibinfo{title}{``Ultrasonic attenuation  peak during fatigue of polycrystalline copper"}}, 
 \bibinfo{journal}{Acta Mater.} 
 \textbf{\bibinfo{volume}{48}}, 
 \bibinfo{pages}{517--524}.
 

\bibitem[{Leisure and Willis(1997)}]{leisure97}  
\bibinfo{author}{Leisure, R. G. and Willis, F. A.}
  (\textbf{\bibinfo{year}{1997}}). 
  {\bibinfo{title}{``Resonant Ultrasound Spectroscopy"}}, 
  \bibinfo{journal}{J. Phys. Condens. Matter} 
  \textbf{\bibinfo{volume}{9}},
   \bibinfo{pages}{6001--6029}.
   
 \bibitem[{Maurel et al.(2004)}]{maureletal}  
\bibinfo{author}{Maurel, A., Mercier, J.-F. and Lund, F.}
  (\textbf{\bibinfo{year}{2004}}). 
  {\bibinfo{title}{``Scattering of an elastic wave by a single dislocation"}}, 
  \bibinfo{journal}{J. Acoust. Soc. Am.} 
  \textbf{\bibinfo{volume}{115}},
   \bibinfo{pages}{2773--2780}.
   
   
 \bibitem[{McSkimin(1961)}]{McSkimin1960}
 \bibinfo{author}{McSkimin, H. J.}
 (\textbf{\bibinfo{year}{1961}}). 
 {\bibinfo{title}{``Pulse Superposition Method for Measuring Ultrasonic Wave Velocities in Solids"}}, 
 \bibinfo{journal}{J. Acoust. Soc. Am.}, 
 \textbf{\bibinfo{volume}{33}}, 
 \bibinfo{pages}{12--16}.
 
\bibitem[{Migliori et. al.(1993)}]{migliori93}  
\bibinfo{author}{Migliori, A., Sarrao, J. L., Visscher, William M., Bell, T. M., Lei, Ming, Fisk, Z. and Leisure, R. G.}
(\textbf{\bibinfo{year}{1993}}). 
{\bibinfo{title}{``Resonant Ultrasound Spectroscopic Techniques for Measurement of the Elastic Moduli of Solids"}}, \bibinfo{journal}{Physica B} \textbf{\bibinfo{volume}{183}},
 \bibinfo{pages}{1--24}.
 
 \bibitem[{Min and Kato(2004)}]{minkato04}
 \bibinfo{author}{Min, X. H. and Kato, H.}
 (\textbf{\bibinfo{year}{2004}}). 
 {\bibinfo{title}{``Change in ultrasonic parameters with loading/unloading process in cyclic loading of aluminium alloy"}}, \bibinfo{journal}{Mater. Sci. Engin.} 
 \textbf{\bibinfo{volume}{A 372}}, 
 \bibinfo{pages}{269--277}.
 
 \bibitem[{Min et al.(2005)}]{minetal}
 \bibinfo{author}{Min, X. H., Kato, H., Narisawa, N. and Kageyama, K.}
 (\textbf{\bibinfo{year}{2005}}). 
 {\bibinfo{title}{``Real-time ultrasonic measurement during tensile testing of aluminum alloys"}}, 
 \bibinfo{journal}{Mater. Sci. Engin.} 
 \textbf{\bibinfo{volume}{A 392}}, 
 \bibinfo{pages}{87--93}.   
 
\bibitem[{Mujica et. al. (2012)}]{Mujica2012}
\bibinfo{author}{Mujica, N., Cerda, M.T., Espinoza, R., Lisoni, J. and Lund, F.}
  (\textbf{\bibinfo{year}{2012}}). 
  {\bibinfo{title}{``Ultrasound as a probe of dislocation density in aluminium"}}, \bibinfo{journal}{Acta Mater.} \textbf{\bibinfo{volume}{60}},
  \bibinfo{pages}{5828--5837}.
  
\bibitem[{Ogi et al.(2002)}]{Ogi2002}
\bibinfo{author}{Ogi, H., Sato, K., Asada, T. and Hirao, M.}
  (\textbf{\bibinfo{year}{2002}}). 
  {\bibinfo{title}{``Complete mode identification for resonance ultrasound spectroscopy"}}, \bibinfo{journal}{J. Acoust. Soc. Am.} \textbf{\bibinfo{volume}{112}},
  \bibinfo{pages}{2553--2557}.
  
\bibitem[{Reed-Hill and Abbaschian(1991)}]{Reedhill1992}
 \bibinfo{author}{Reed-Hill, R.E. and Abbaschian, R.}
 (\textbf{\bibinfo{year}{1991}}). 
 {\bibinfo{booktitle}{``Physical Metallurgy Principles"}}, 
 \bibinfo{publisher}{(WS Publishing Company, Boston)}.

\end{thebibliography}


\end{document}